\documentclass[twocolumn]{aastex6}

\shorttitle{Solar Dynamo in Transition}
\shortauthors{Metcalfe et al.}
\slugcomment{Accepted by the Astrophysical Journal Letters}

\begin{document}

\title{Stellar Evidence that the Solar Dynamo may be in Transition}

\author{Travis S.\ Metcalfe$^{1,2}$, Ricky Egeland$^{3,4}$, Jennifer van Saders$^{5,6}$}

\affil{$^1${Space Science Institute, 4750 Walnut St.\ Suite 205, Boulder CO 80301 USA}\\
$^2${Visiting Scientist, National Solar Observatory, 3665 Discovery Dr., Boulder CO 80303 USA}\\
$^3${High Altitude Observatory, National Center for Atmospheric Research, P.O. Box 3000, Boulder CO 80307 USA}\\
$^4${Department of Physics, Montana State University, Bozeman MT 59717 USA}\\
$^5${Carnegie Observatories, 813 Santa Barbara St, Pasadena CA 91101 USA}\\
$^6${Department of Astrophysical Sciences, Princeton University, Princeton NJ 08544 USA}}

\begin{abstract}

Precise photometry from the {\it Kepler} space telescope allows not only 
the measurement of rotation in solar-type field stars, but also the 
determination of reliable masses and ages from asteroseismology. These 
critical data have recently provided the first opportunity to calibrate 
rotation-age relations for stars older than the Sun. The evolutionary 
picture that emerges is surprising: beyond middle-age the efficiency of 
magnetic braking is dramatically reduced, implying a fundamental change in 
angular momentum loss beyond a critical Rossby number (Ro\,$\sim$\,2). We 
compile published chromospheric activity measurements for the sample of 
{\it Kepler} asteroseismic targets that were used to establish the new 
rotation-age relations. We use these data along with a sample of 
well-characterized solar analogs from the Mount Wilson HK survey to 
develop a qualitative scenario connecting the evolution of chromospheric 
activity to a fundamental shift in the character of differential rotation. 
We conclude that the Sun may be in a transitional evolutionary phase, and 
that its magnetic cycle might represent a special case of stellar dynamo 
theory.\\

\end{abstract}


\section{Introduction}

One of the biggest surprises from the {\it Kepler} space telescope is that 
there were so few surprises for stars near the main-sequence. Stellar 
evolution models that had been calibrated to match the properties of the 
Sun were generally sufficient to reproduce asteroseismic observations of 
other solar-type stars. For example, \cite{Metcalfe2015} used a solar data 
set comparable to what {\it Kepler} obtained for other solar-type stars 
and recovered the solar age and other properties with an absolute accuracy 
better than 5\%. Identical methods applied independently to the complete 
{\it Kepler} data sets for 16~Cyg~A~\&~B found the same age and 
composition for the two components, bolstering our confidence in the 
reliability of asteroseismic techniques.

A major surprise for {\it Kepler} main-sequence stars arose when 
asteroseismic properties were confronted with measurements of rotation. 
The idea of using rotation as a diagnostic of stellar age dates back to 
\cite{Skumanich1972}, and decades of effort have gone into calibrating the 
modern concept of {\it gyrochronology} \citep{Barnes2007}. The initial 
contributions from {\it Kepler} included observations of stellar rotation 
in the 1~Gyr-old cluster NGC~6811 \citep{Meibom2011} and the 2.5~Gyr-old 
cluster NGC~6819 \citep{Meibom2015}, extending the calibration of the 
method significantly beyond previous work. Initial indications of a 
possible conflict between asteroseismology and gyrochronology were noted 
by \cite{Angus2015}, who found that no single color-dependent relationship 
between rotation and age could simultaneously describe the cluster and 
field populations. Although they used low-precision asteroseismic ages 
from grid-based modeling \citep{Chaplin2014}, the tension was still 
evident.

\floattable
\begin{deluxetable}{lccccccr}
\tablecaption{Properties of the asteroseismic sample with rotation and activity data.\label{tab1}}
\tablehead{\colhead{KIC} & \colhead{\bv} & \colhead{$R/R_\odot$} & \colhead{$M/M_\odot$} & \colhead{$t$/Gyr} & 
\colhead{$P_{\rm rot}$} & \colhead{$\log R^\prime_{HK}$} & \colhead{Sources}}
\startdata
 3656476 & 0.782 & $1.335\pm0.009$ & $1.13\pm0.03$ & $8.67\pm0.61$ & $31.67\pm3.53$       & $-$5.109 & 1,6,11 \\
 5184732 & 0.696 & $1.367\pm0.013$ & $1.27\pm0.04$ & $4.17\pm0.39$ & $19.79\pm2.43$       & $-$5.130 & 1,6,8  \\
 6116048 & 0.589 & $1.219\pm0.009$ & $1.01\pm0.03$ & $6.23\pm0.37$ & $17.26\pm1.96$       & $-$5.019 & 2,6,9  \\
 6521045 & 0.790 & $1.474\pm0.013$ & $1.04\pm0.02$ & $6.24\pm0.64$ & $25.34\pm2.78$       & $-$5.042 & 4,4,10 \\
 7680114 & 0.690 & $1.421\pm0.012$ & $1.12\pm0.03$ & $7.35\pm0.78$ & $26.31\pm1.86$       & $-$4.905 & 1,6,11 \\
 8006161 & 0.859 & $0.947\pm0.007$ & $1.04\pm0.02$ & $5.04\pm0.17$ & $29.79\pm3.09$       & $-$5.011 & 2,6,9  \\
 8379927 & 0.570 & $1.130\pm0.013$ & $1.15\pm0.04$ & $1.67\pm0.12$ & $16.99\pm1.35$       & $-$4.834 & 5,6,9  \\
 9098294 & 0.567 & $1.154\pm0.009$ & $1.00\pm0.03$ & $7.28\pm0.51$ & $19.79\pm1.33$       & $-$5.020 & 2,6,9  \\
 9139151 & 0.520 & $1.146\pm0.011$ & $1.14\pm0.03$ & $1.71\pm0.19$ & $10.96\pm2.22$       & $-$4.954 & 2,6,9  \\
 9955598 & 0.713 & $0.907\pm0.005$ & $0.96\pm0.03$ & $6.43\pm0.40$ & $34.20\pm5.64$       & $-$5.048 & 4,6,10 \\
10454113 & 0.528 & $1.250\pm0.015$ & $1.19\pm0.04$ & $2.03\pm0.29$ & $14.61\pm1.09$       & $-$4.872 & 2,6,9  \\
10644253 & 0.598 & $1.108\pm0.016$ & $1.13\pm0.05$ & $1.07\pm0.25$ & $10.91\pm0.87$       & $-$4.696 & 2,6,11 \\
10963065 & 0.509 & $1.222\pm0.010$ & $1.07\pm0.03$ & $4.36\pm0.29$ & $12.58\pm1.70$       & $-$5.054 & 4,6,10 \\
11244118 & 0.732 & $1.589\pm0.026$ & $1.10\pm0.05$ & $6.43\pm0.58$ & $23.17\pm3.89$       & $-$5.148 & 2,6,9  \\
12009504 & 0.556 & $1.375\pm0.015$ & $1.12\pm0.03$ & $3.64\pm0.26$ & $9.39\pm0.68$        & $-$4.977 & 2,6,9  \\
12069424 & 0.643 & $1.236\pm0.008$ & $1.10\pm0.02$ & $7.07\pm0.46$ & $23.8^{+1.5}_{-1.8}$ & $-$5.105 & 3,3,7  \\
12069449 & 0.661 & $1.123\pm0.007$ & $1.06\pm0.02$ & $6.82\pm0.28$ & $23.2^{+11.5}_{-3.2}$& $-$5.094 & 3,3,7  \\
12258514 & 0.599 & $1.573\pm0.010$ & $1.19\pm0.03$ & $4.03\pm0.32$ & $15.00\pm1.84$       & $-$5.024 & 2,5,9  \\
\enddata
\tablerefs{(1)~\citet{Mathur2012}; (2)~\citet{Metcalfe2014}; (3)~\citet{Davies2015};
(4)~\citet{Ceillier2016}; (5)~\citeauthor{Davies2015}~(in preparation); (6)~\citet{Garcia2014};
(7)~\citet{Wright2004}; (8)~\citet{Isaacson2010}; (9)~\citet{Karoff2013};
(10)~\citet{Marcy2014}; (11)~\citet{Salabert2016}}
\vspace*{-24pt}
\end{deluxetable}

The source of disagreement between the age scales from asteroseismology 
and gyrochronology came into focus after \cite{vanSaders2016} scrutinized 
{\it Kepler} targets with precise ages from detailed modeling. They 
confirmed the existence of a population of field stars rotating more 
quickly than expected from gyrochronology. They discovered that the 
anomalous rotation became significant near the solar age for G-type 
targets, but it appeared earlier in F-type stars and later in K-type 
stars. This dependence on spectral type suggested a connection to the 
Rossby number (Ro $\equiv P_{\rm rot}/\tau_c$), the ratio of the rotation 
period to the convective turnover time. They proposed that magnetic 
braking operates with a dramatically reduced efficiency beyond a critical 
Rossby number, and they reproduced the observations with rotational 
evolution models that eliminated angular momentum loss beyond 
Ro\,$\sim$\,2.

Motivated by these results, we search for a magnetic counterpart to the 
rotational transition discovered by \cite{vanSaders2016}. We compile 
published chromospheric activity measurements for the {\it Kepler} sample, 
and we compare them to a selection of G-type stars from the Mount Wilson 
HK survey \citep{Baliunas1996,Donahue1996}. We use these observations to 
reinterpret some well-established features of chromospheric activity in 
solar-type stars, and we propose a qualitative scenario for magnetic 
evolution that connects the available evidence to a shift in the character 
of differential rotation (section~\ref{SEC2}). We then use a sample of 
well-characterized solar analogs to constrain the evolution of key dynamo 
ingredients and outputs (section~\ref{SEC3}), with a particular focus on 
the magnetic topology revealed by Zeeman Doppler imaging 
\citep[ZDI,][]{Petit2008}. We conclude that the Sun may be in a 
transitional evolutionary phase, and we discuss the implications for 
understanding the solar magnetic cycle in the context of activity cycles 
observed in other stars (section~\ref{SEC4}).

\section{Rotation-activity relation \label{SEC2}}

We have identified chromospheric activity measurements for 18 of the 21 
stars in the \cite{vanSaders2016} sample, which are listed with their 
asteroseismic properties and rotation periods in Table~\ref{tab1}. All of 
the asteroseismic data were analyzed using the method described in 
\cite{Metcalfe2014}, and most of the rotation periods were obtained from 
the method described in \cite{Garcia2014}. The activity data come from 
multiple sources, but we gave priority to measurements made by 
\cite{Karoff2013} during the {\it Kepler} mission.

Our sample is dominated by stars more evolved than the Sun because of two 
selection effects. First, the intrinsic amplitudes of solar-like 
oscillations scale with the ratio of luminosity to mass \citep{Houdek1999, 
Samadi2001}, making them easier to detect in early-type and more evolved 
stars. Second, the observed amplitudes are suppressed by strong magnetic 
activity \citep{Chaplin2011}, biasing detections away from young 
main-sequence stars. Consequently, the youngest and most active targets in 
our sample are F-type stars.

\begin{figure*}
\centerline{\includegraphics[angle=270,width=6.25in]{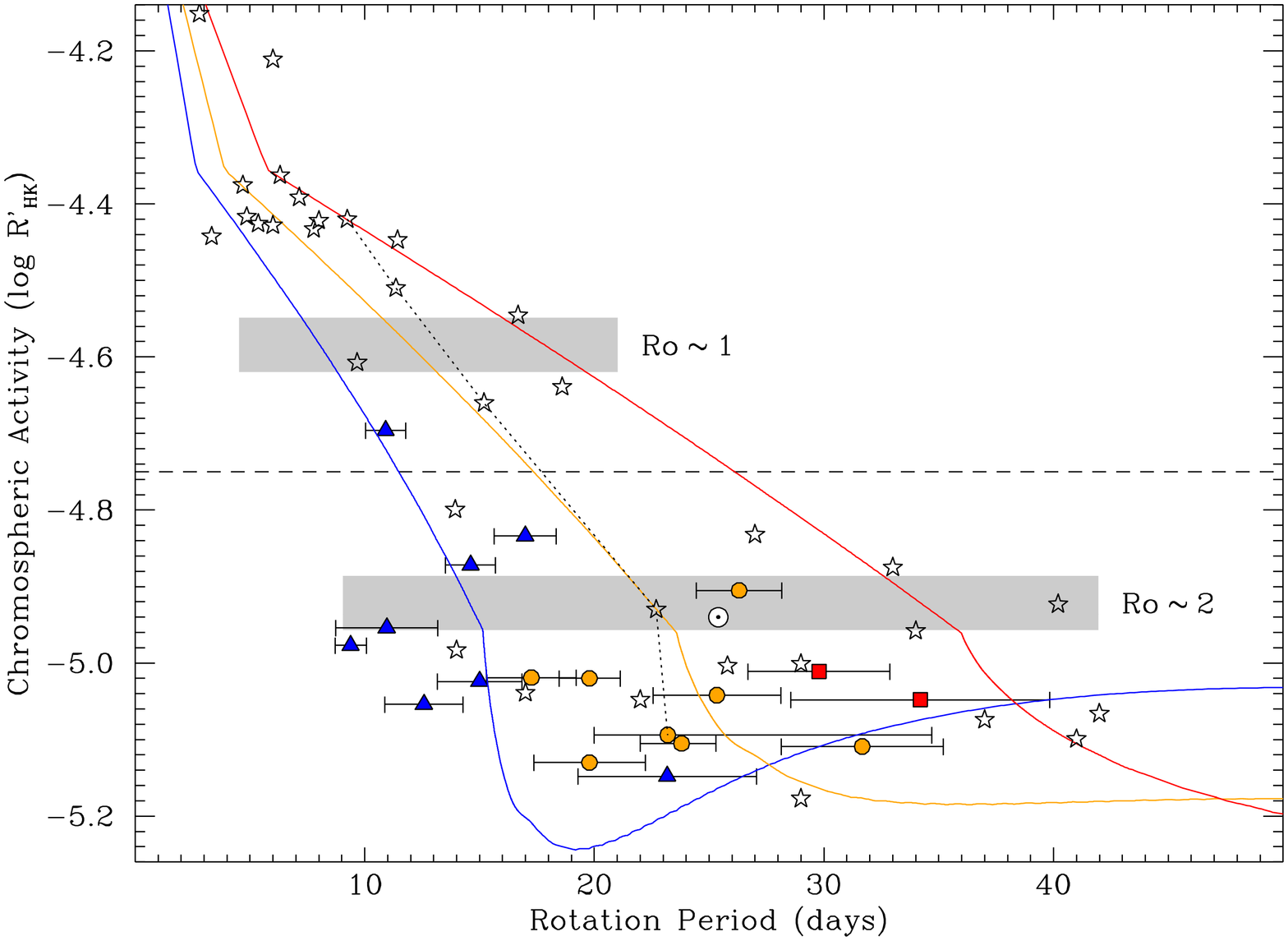}}
\caption{Relationship between chromospheric activity and rotation in 
field dwarfs and subgiants. G-type stars from the Mount Wilson HK project 
are shown as star symbols. The asteroseismic sample from {\it Kepler} is 
shown as colored points, including F-type (blue triangles), G-type (orange 
circles), and K-type (red squares) stars. The Sun is shown with its usual 
symbol ($\odot$). The dashed line indicates the activity level of the 
Vaughan-Preston gap while the dotted line connects several 
well-characterized solar analogs. Shaded regions denote the activity 
levels that correspond to key Rossby numbers, using the empirical 
activity-rotation relation from \cite{Mamajek2008}. Colored curves use 
this same relation to plot rotational evolution models from 
\cite{vanSaders2016} for KIC\,10963065 (blue), KIC\,6521045 (orange), and 
KIC\,9955598 (red).\label{fig1}}
\end{figure*}

The relationship between chromospheric activity and rotation for our 
sample is illustrated in Figure~\ref{fig1} with colored points, including 
F-type (blue triangles), G-type (orange circles), and K-type (red squares) 
targets. For context, we also show a selection of G-type stars from the 
Mount Wilson HK survey (star symbols) and a few rotational evolution 
models from \cite{vanSaders2016}, converted from Rossby number to 
chromospheric activity using the rotation-activity relation of 
\cite{Mamajek2008}. We extrapolate this relation for $\log R^{\prime}_{\rm 
HK}<-5.0$, where the evolution tracks exceed a critical Rossby number 
(Ro$_{crit}$\,=\,2.1) and magnetic braking ceases. The activity levels 
that correspond to key Rossby numbers are shown as shaded regions on 
either side of the Vaughan-Preston gap \citep[dashed 
line,][]{Vaughan1980}. The vertical extent of the shaded regions indicates 
the variation of solar activity through a magnetic cycle \citep{Hall2004}, 
while the horizontal extent corresponds to the variation in convective 
turnover times for the color range of our sample. The dotted line connects 
several well-characterized solar analogs (see section~\ref{SEC3}).

The horizontal spread in rotation rates at low activity levels is evident 
both in our sample and in the Mount Wilson stars. This can largely be 
understood as a consequence of how the convective turnover time depends on 
spectral type, such that stars with deeper convection zones spend more 
time at a given activity level. At the lowest activity levels the 
evolution changes abruptly, leading to an excess of evolved stars with 
relatively short rotation periods. This can most easily be seen for the 
F-type and G-type stars, with rotation periods around 10 and 20 days 
respectively.

\floattable
\begin{deluxetable}{lccccc}
\tablecaption{Properties of the solar analog sample.\label{tab2}}
\tablehead{\colhead{~} & \colhead{HD\,20630} & \colhead{HD\,30495} & \colhead{HD\,76151} & \colhead{18~Sco} & \colhead{16~Cyg~B} } 
\startdata
Spectral Type          & G5V             & G1.5V           & G3V             & G2V                    & G3V             \\
$R/R_\odot$            & $0.917\pm0.011$ & $0.983\pm0.016$ & $0.979\pm0.017$ & $1.010\pm0.009$        & $1.116\pm0.006$ \\
$M/M_\odot$            & $0.95\pm0.09$   & $0.86\pm0.12$   & $1.24\pm0.12$   & $1.03\pm0.01$          & $1.04\pm0.02$   \\
Age~[Gyr]              & $0.5\pm0.1$     & $1.0\pm0.1$     & $1.4\pm0.2$     & $3.66^{+0.44}_{-0.50}$ & $6.74\pm0.24$   \\
$P_{\rm rot}$~[d]      & 9.24            & 11.36           & 15.2            & 22.7                   & 23.2            \\
$\Delta P/P$           & 0.0509          & 0.0519          & $\cdots$        & $\cdots$               & $\cdots$        \\
Ro                     & 0.70            & 1.08            & 1.18            & 1.92                   & 2.10            \\
$\log R^\prime_{\rm HK}$&$-$4.42         & $-$4.51         & $-$4.66         & $-$4.93                & $-$5.09         \\
$P_{\rm cyc}$(A)~[yr]  & 5.6             & 12.2            & $\cdots$        & $\cdots$               & $\cdots$        \\
$P_{\rm cyc}$(I)~[yr]  & $\cdots$        & 1.67            & 2.52            & 7.1                    & $\cdots$        \\
Sources                & 1,5,7,11,14     & 1,6,11          & 1,5,8,11,14     & 2,4,9,12               & 3,10,13         \\
\enddata
\tablerefs{(1)~\citet{Valenti2005}; (2)~\citet{Bazot2011}; (3)~\citet{Metcalfe2015}; (4)~\citet{Li2012}; (5)~\citet{Barnes2007}; 
(6)~\citet{Egeland2015}; (7)~\citet{Donahue1996}; (8)~\citet{Olah2016}; (9)~\citet{Petit2008}; (10)~\citet{Davies2015}; 
(11)~\citet{Baliunas1996}; (12)~\citet{Hall2007a}; (13)~\citet{Wright2004}; (14)~\citet{Baliunas1995}}
\vspace*{-12pt}
\end{deluxetable}

As suggested by \cite{vanSaders2016}, one way to reduce the efficiency of 
magnetic braking is to concentrate the field into smaller spatial scales. 
For example, \cite{Reville2015} demonstrated that the dipole component of 
the field is responsible for most of the angular momentum loss due to the 
magnetized stellar wind. {\it We propose that a change in the character of 
differential rotation is the underlying mechanism that ultimately disrupts 
the large-scale organization of magnetic field in solar-type stars.} The 
process begins at Ro\,$\sim$\,1, where the rotation period becomes 
comparable to the convective turnover time. Differential rotation is an 
emergent property of turbulent convection in the presence of Coriolis 
forces, and \cite{Gastine2014} showed that many global convection 
simulations exhibit a transition from solar-like to anti-solar 
differential rotation near Ro\,$\sim$\,1.

We can then interpret the Vaughan-Preston gap as a signature of rapid 
magnetic evolution triggered by a shift in the character of differential 
rotation. \cite{Pace2009} used activity measurements of stars in several 
open clusters to constrain the age of F-type stars crossing the gap to be 
between 1.2 and 1.4~Gyr. The two most active F-type stars in our sample 
have ages of 1.07 and 1.67~Gyr and fall on opposite sides of the gap, 
consistent with the results of \cite{Pace2009} and validating our 
asteroseismic age scale.

According to \cite{Lockwood2007}, the Vaughan-Preston gap also corresponds 
to a shift from spots to faculae as the dominant source of photometric 
variability on the timescale of stellar cycles. \cite{Shapiro2014} noted 
that the surface area of spots varies quadratically with activity level 
over the solar cycle, while faculae vary linearly. Extrapolating to higher 
activity levels they reproduced the transition from spot- to 
faculae-dominated variability across the Vaughan-Preston gap. They do not 
offer an explanation for the quadratic dependence of spot area on activity 
level in the Sun. We speculate that an accelerated decrease in the surface 
area of spots may be one consequence of a disruption of differential 
rotation at Ro\,$\sim$\,1.

Emerging from the rapid magnetic evolution across the Vaughan-Preston gap, 
stars reach the Ro\,$\sim$\,2 threshold where magnetic braking operates 
with a dramatically reduced efficiency, possibly due to a shift in 
magnetic topology (see section~\ref{SEC3}). The rotation period then 
evolves as the star undergoes slow expansion and changes its moment of 
inertia as it ages. At the same time, the activity level decreases with 
effective temperature as the star expands and mechanical energy from 
convection largely replaces magnetic energy driven by rotation as the 
dominant source of chromospheric heating \citep{BohmVitense2007}.

\section{Evidence from solar analogs \label{SEC3}}

Although the basic stellar properties of the {\it Kepler} sample are 
well-constrained from asteroseismology, characterization of their magnetic 
cycles and differential rotation will require additional observations and 
analysis. In the meantime, we use a selection of well-characterized solar 
analogs to investigate the evolution of key dynamo ingredients and outputs 
for solar-type stars as they pass through various stages of the scenario 
outlined in section~\ref{SEC2}. These solar analogs are connected with a 
dotted line in Figure~\ref{fig1}, their properties are listed in 
Table~\ref{tab2}, and we discuss them below from high to low activity.

The most active of our selected solar analogs is HD\,20630 
($\kappa^1$~Cet), which has not yet experienced a putative shift in the 
character of differential rotation at Ro\,$\sim$\,1. \cite{Walker2007} 
modeled three seasons of photometry from the MOST satellite to establish 
that HD\,20630 shows solar-like differential rotation, as expected. The 
latitudinal shear is slightly smaller than in the Sun, but the modeled 
spots appear over a wide range of active latitudes from $10^\circ$ to 
$75^\circ$. \cite{Baliunas1995} identified a 5.6 year cycle in the 
chromospheric activity. For a rotation period of 9.24~d 
\citep{Donahue1996}, this is a normal cycle on the active (A) branch 
\citep{SB1999,BohmVitense2007}.

Slightly closer to the Ro\,$\sim$\,1 transition is HD\,30495, which was 
recently characterized by \cite{Egeland2015} from 47 years of activity 
measurements and 22 years of photometry. The average rotation period of 
11.36~d varied between seasons, suggesting a broad range of active 
latitudes and implying a surface shear less than or comparable to the Sun. 
The unknown latitudes of the spots prevented the authors from determining 
the sense of the differential rotation. However, the brightness variations 
were unambiguously dominated by spots rather than faculae, as expected for 
a star on the active side of the Vaughan-Preston gap. They identified 
magnetic variability on two different timescales, with an average long 
cycle of 12.2 years and persistent short-term variability around 1.67 
years. These periods are comparable to the sunspot cycle and 
quasi-biennial variations \citep{Bazilevskaya2014} but in a star rotating 
at more than twice the rate of the Sun. The long cycle falls squarely on 
the A-branch, while the short-period variations can be interpreted as a 
secondary cycle on the inactive (I) branch \citep{SB1999,BohmVitense2007}.

Just above the activity level of the Vaughan-Preston gap is HD\,76151, one 
of the solar analogs observed with ZDI by \cite{Petit2008}. The equatorial 
rotation period inferred from spectropolarimetry (20.5~d) is significantly 
longer than that determined from chromospheric activity measurements 
\citep[15.2~d,][]{Olah2016}, which reflects rotation at the active 
latitudes. If both periods are reliable, it implies strong anti-solar 
differential rotation in this star. \cite{Petit2008} found no indication 
of surface differential rotation, but they only made such a detection for 
the most active star in their sample. However, they found that 93\% of the 
magnetic energy in HD\,76151 was in the poloidal component of the field. 
More generally, they found that a growing fraction of the field was 
poloidal in solar analogs as rotation slowed. In flux-transport dynamo 
models, differential rotation is the mechanism that recycles poloidal 
field into the toroidal component \citep[e.g.,][]{Dikpati2009}, so this 
trend can be interpreted as evidence of inefficient differential rotation. 
Decomposing the poloidal field into spherical harmonics, they found 79\% 
in the dipole component and 18\% in the quadrupole. HD\,76151 has the 
shortest activity cycle reported by the Mount Wilson HK survey 
\citep[2.52~years,][]{Baliunas1995}, which falls on the I-branch.

On the opposite side of the Vaughan-Preston gap near Ro\,$\sim$\,2 is the 
well-known solar twin HD\,146233 (18~Sco). As expected, \cite{Hall2007b} 
found the brightness variations to be dominated by faculae just as in the 
Sun. \cite{Petit2008} found more than 99\% of the magnetic energy 
concentrated in poloidal field, a further indication of the diminishing 
role of differential rotation. Most interesting from the standpoint of 
weakened magnetic braking, only 34\% of the poloidal field was in the 
dipole component while 56\% was in the quadrupole. The ZDI measurements 
were obtained near a maximum in magnetic activity, so we might expect this 
solar twin to be dominated by the quadrupole. Indeed, a low-order 
reconstruction of the solar magnetic field near maximum by 
\cite{Vidotto2016} resembles the ZDI map of 18~Sco. \cite{Hall2007b} 
identified a 7 year activity cycle in 18~Sco, which appears to be a normal 
cycle on the I-branch unlike the 11 year solar cycle.

Beyond the transition to reduced magnetic braking is the solar analog 
binary 16~Cyg~A~\&~B, which is also part of our {\it Kepler} sample. 
\cite{Garcia2014} found no significant photometric variation, so the 
rotation periods were determined by \cite{Davies2015} from frequency 
splitting of the oscillation modes. For a sample of {\it Kepler} targets 
that also showed spot modulations, \cite{Nielsen2015} demonstrated that 
the two methods yield consistent results. They also noted an absence of 
significant radial differential rotation in their sample, providing 
further evidence of minimal shear at these low activity levels. Despite an 
impressive signal-to-noise ratio, P.~Petit (priv.~comm.) found no 
detectable Zeeman signatures in spectropolarimetry of 16~Cyg~A~\&~B. If 
present, the dipole component would be the easiest signal to detect. 
Chromospheric activity monitoring from Lowell Observatory since 1994 
\citep{Hall2007a} shows both components with activity levels below that of 
the Sun at solar minimum, and no cyclic variation.

\begin{figure*}
\centerline{\includegraphics[angle=270,width=6.15in]{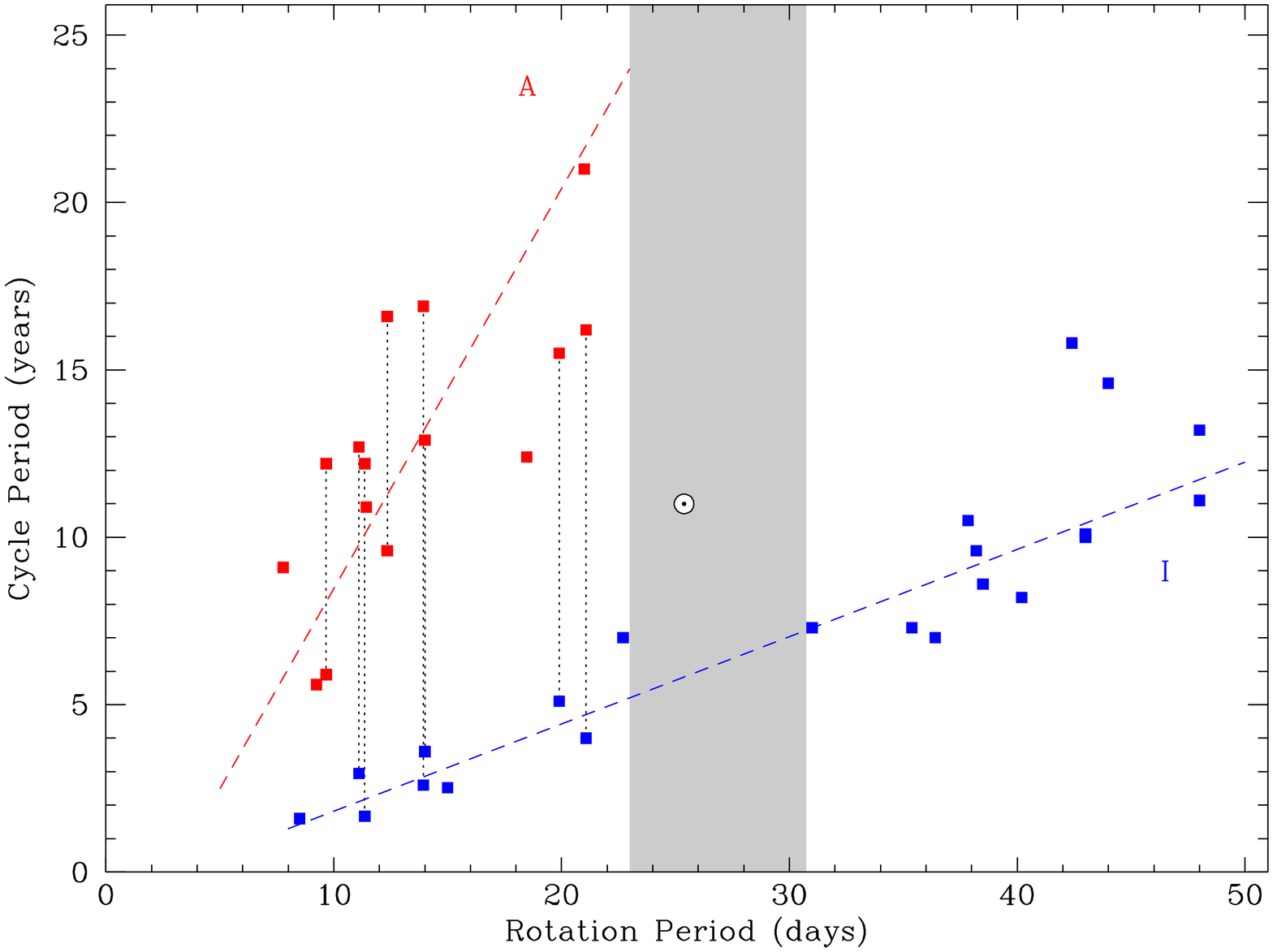}}
\caption{Updated version of a diagram published in \cite{BohmVitense2007} 
using data from \cite{SB1999}, showing the active (A) and inactive (I) 
branches. More recent data have been added from \cite{Hall2007b}, 
\cite{Metcalfe2010,Metcalfe2013}, and \cite{Egeland2015}. Multiple cycles 
observed in the same star are connected with vertical dotted lines. The 
shaded region indicates the range of rotation periods around the Sun 
($\odot$) where the Mount Wilson targets are flat-activity 
stars.\label{fig2}}
\vspace*{-1pt}
\end{figure*}

\section{Conclusions \& Discussion \label{SEC4}}

Based on the scenario for magnetic evolution outlined in 
section~\ref{SEC2} and considering the evidence from solar analogs 
presented in section~\ref{SEC3}, we conclude that the Sun may be in a 
transitional evolutionary phase and that it might represent a special case 
of stellar dynamo theory. The Sun obviously exhibits solar-like 
differential rotation, so the transition at Ro\,$\sim$\,1 suggested by 
simulations may be more complex than a shift to anti-solar latitudinal 
shear. The Sun appears to have a narrower range of active latitudes than 
the most active solar analogs, and the torsional oscillations observed 
from helioseismology \citep{Vorontsov2002} may be one indication of an 
ongoing transition in the character of differential rotation. The 
concentration of magnetic energy into poloidal field for solar analogs 
that span the Vaughan-Preston gap is further evidence of relatively 
inefficient differential rotation. The Sun still exhibits a dipole 
component to its global field, particularly near magnetic minimum, but the 
solar analogs also suggest a gradual concentration of the field into 
smaller spatial scales, leading to weakened magnetic braking. Such a 
transition may represent the shift from a dominant $\alpha$-$\Omega$ 
dynamo to an $\alpha^2$ dynamo, but this remains to be demonstrated.

Perhaps the most significant indication that the Sun is in a transitional 
evolutionary phase is its 11 year magnetic cycle, which falls between the 
active and inactive branches established by other stars 
(see Figure~\ref{fig2}). Even the slightly younger solar twin 18~Sco 
exhibits a normal cycle on the I-branch, while the Sun is the only star 
with a rotation period between 23-30~d that shows a cycle at all. There 
were several Mount Wilson targets with rotation periods in this range, but 
they are all flat-activity stars. The Maunder minimum may represent one 
manifestation of the Sun beginning to enter this flat-activity phase. The 
fact that all of the slower rotators with cycles are K-type stars is 
perfectly understandable, since magnetic braking ceases in G-type stars 
before they reach these long rotation periods. Among the faster rotators, 
many of the Mount Wilson targets appeared to have chaotic variability in 
their chromospheric activity. This may be due to the ubiquity of secondary 
cycles on the I-branch, combined with seasonal data gaps that fail to 
sample these short periods adequately.

Future observations and analysis will offer additional tests of the 
scenario for magnetic evolution described in section~\ref{SEC2}. The 
sample of {\it Kepler} targets with asteroseismic ages and detected 
rotation periods promises to expand to $\sim$30 stars from analysis of the 
full-length data sets (Lund et al.~in preparation). Only a few of these 
targets have not yet made the transition to weakened magnetic braking, so 
further characterization of the bright solar analogs will be essential to 
establish the nature and timing of the chain of events that lead to this 
evolutionary phase. Additional ZDI measurements of solar analogs, more 
constraints on differential rotation at various activity levels, and 
asteroseismic observations of Mount Wilson stars with the K2 and TESS 
missions will all be helpful.
\vspace*{-4pt}

\acknowledgements The authors would like to thank Axel Brandenburg, Paul 
Charbonneau, Mark Giampapa, Peter Gilman, Marc Pinsonneault, Steve Saar, 
Dave Soderblom and Ellen Zweibel for helpful discussions. This work was 
supported in part by NASA grant NNX15AF13G. R.~E. is supported by a 
Newkirk Fellowship at the High Altitude Observatory.

\bibliographystyle{aasjournal}

\end{document}